\documentclass[twocolumn,aps]{revtex4}

\usepackage{graphicx}% Include figure files
\usepackage{dcolumn}% Aligntable columns on decimal point
\usepackage{bm}% bold math
\usepackage{times}

\usepackage{color}
\usepackage{amstext}

\begin{document}

\title{Charge transport and magnetization profile at the interface between a correlated metal \newline and an antiferromagnetic insulator}

\author{J.W.~Freeland$^1$, J.~Chakhalian$^2$, A.V.~Boris$^{3,4}$, J.-M.~Tonnerre$^{5}$, J.J.~Kavich$^{1,6}$, P.~Yordanov$^3$, S. Grenier$^5$, P.~Popovich$^3$, H.N.~Lee$^7$, B.~Keimer$^3$}

\affiliation{$^1$Advanced Photon Source, Argonne National Laboratory, Argonne, IL 60439, USA}
\affiliation{$^2$Department of Physics, University of Arkansas, Fayetteville, AR 72701, USA}
\affiliation{$^3$Max-Planck-Institut f\"{u}r Festk\"{o}rperforschung, D-70569 Stuttgart, Germany}
\affiliation{$^4$Department of Physics, Loughborough University, Loughborough, LE11 3TU, United Kingdom}
\affiliation{$^5$Institut N\'eel, CNRS and Universit\'e Joseph Fourier, Boite Postale 166, F-38043 Grenoble Cedex 9, France}
\affiliation{$^6$Department of Physics, University of Illinois at Chicago, Chicago, IL 60607, USA}
\affiliation{$^7$Materials Science and Technology Division, Oak Ridge National Laboratory, Oak Ridge, TN 37831, USA}

\begin{abstract}
A combination of spectroscopic probes was used to develop a detailed
experimental description of the transport and magnetic properties of
superlattices composed of the paramagnetic metal CaRuO$_3$ and the
antiferromagnetic insulator CaMnO$_3$. The charge carrier
density and Ru valence state in the superlattices are not
significantly different from those of bulk CaRuO$_3$. The small
charge transfer across the interface implied by these observations
confirms predictions derived from density functional calculations.
However, a ferromagnetic polarization due to canted Mn spins
penetrates 3-4 unit cells into CaMnO$_3$, far exceeding the
corresponding predictions. The discrepancy may indicate the
formation of magnetic polarons at the interface.
\end{abstract}

\maketitle

Recent advances in the atomic-scale control of transition metal
oxide interfaces offer new opportunities for the manipulation of
strongly correlated electron systems. While first device
applications are being explored \cite{ahn06}, there is growing
awareness of the wealth of microscopic phenomena that need to be
understood in order to arrive at a quantitative description of the
electronic state at oxide interfaces. These include the polarity of
the atomic layers comprising the interface \cite{nakagawa}, lattice
relaxations (which are particularly important for ferroelectric and
quantum paraelectric materials) \cite{pentcheva}, orbital
reconstructions induced by misfit strain \cite{tebano} and/or
covalent bonding across the interface \cite{chakhalian1}, magnetic
polarization due to the reduced coordination of magnetic atoms
and/or exchange interactions across the interface
\cite{chakhalian2}, and interface-specific disorder. This calls for
simple model systems in which individual issues can be isolated, and
the results of theoretical and experimental studies of buried
interfaces can be quantitatively compared. A model system that has
received particular recent attention is the weakly correlated
metallic state at the interface between the band insulators
SrTiO$_3$ and LaAlO$_3$, whose properties are controlled by
interface polarity and lattice relaxations
although there is still an ongoing debate about the role of defects (see Ref.\ \onlinecite{huijbenrev} and references therein).

Another problem of fundamental interest is the interface between a
correlated metallic state and an antiferroagnetic insulator, two of the most
extensively investigated ground states of bulk transition metal
oxides. As theoretical research is beginning to address the length
scales for the intermixing of metallicity and antiferromagnetic
order in this situation \cite{rosch,nanda}, the interface between
the correlated paramagnetic metal CaRuO$_3$ (Refs.
\onlinecite{kamal,lee,mazin}) and the antiferromagnetic
insulator CaMnO$_3$ (Ref. \onlinecite{neumeier}) is emerging as a
model system where these issues can be explored experimentally
without interference from interface polarity, incipient
ferroelectricity, extensive disorder or misfit strain, and orbital
degeneracy. Whereas bulk CaMnO$_3$ is a $G$-type antiferromagnet
(that is, the Mn spin orientation alternates along all
nearest-neighbor bond directions of the nearly cubic perovskite
lattice), prior magnetometry \cite{takahashi} and magneto-optical
\cite{yamada} studies of CaRuO$_3$-CaMnO$_3$ superlattices revealed
a ferromagnetic moment centered at the interface. Based on density
functional calculations, this was attributed to leakage of a small
number of itinerant electrons from CaRuO$_3$ into the first atomic
layer of CaMnO$_3$, where they actuate a ferromagnetic double
exchange interaction and induce canting of the Mn spins
\cite{nanda}.

In this Letter we combine several spectroscopic probes of
CaRuO$_3$-CaMnO$_3$ superlattices including far-infrared (FIR)
spectral ellipsometry, x-ray absorption spectroscopy (XAS), x-ray
magnetic circular dichroism (XCMD), and x-ray resonant magnetic
scattering (XRMS) in order to develop a comprehensive experimental
description of the charge transport properties as well as the
valence state and magnetic polarization of Ru and Mn atoms near the
interface. We find that the charge transfer across the interface is
small, and that the net magnetization of the superlattices arises
mostly from canted Mn spins near the interface, confirming
corresponding predictions of the density functional calculations
\cite{nanda}. Surprisingly, however, the penetration depth of the
ferromagnetic polarization in CaMnO$_3$ is 3-4 unit cells (u.c.),
greatly exceeding the theoretically predicted length scale of 1 u.c.
Based on an analogy to lightly doped bulk CaMnO$_3$, we attribute
the large ferromagnetic penetration depth to polaronic effects not
included in the calculations.

A series of superlattices, LaAlO$_3$ (001)/[CaMnO$_3$(10 u.c.)/CaRuO$_3$(\textit{N} u.c.)]$_6$, with $N=4$ to 10 consecutive CaRuO$_3$ unit cells were grown by pulsed laser deposition
with a KrF excimer laser ($\lambda$ = 248 nm, \textit{J}=1 J/cm$^2$)
at a substrate temperature of 720$^{\circ}$C in 10 mTorr of pure
oxygen, using stoichiometric sintered targets. Upon deposition, all
samples were {\it in-situ} annealed in 50-100 mTorr of oxygen for
5-10 minutes, and then gradually cooled down to room temperature.
Reflection high-energy electron diffraction (RHEED) oscillations
observed {\it in situ} as well as spectroscopic data and XRMS
profiles taken on the finished specimens demonstrate
flat, atomically sharp interfaces. In agreement with prior work on
this system \cite{takahashi,yamada}, magnetic susceptibility
measurements on both samples indicate a net magnetization upon
cooling below 120 K, coincident with the N\'{e}el temperature of
bulk antiferromagnetic CaMnO$_3$. The FIR experiments were conducted
at beamline IR1 at the ANKA synchrotron in Karlsruhe, Germany,
\cite{bernhard} with angles of incidence ranging from 70 to
85$^\circ$. The XAS and XMCD measurements were performed at beamline
4-ID-C of the Advanced Photon Source, with magnetic
fields of up to 7 T applied in the plane of
the superlattices. The measurements were performed by monitoring the
intensity, $I$, of left ($-$) and right ($+$) circularly polarized
x-rays absorbed by the specimen using the surface/interface sensitive total-electron-yield detection described previously
\cite{chakhalian2,chakhalian1}.
The sum, $I^{+}$+$I^{-}$, yields the XAS signal, while the XMCD
signal is derived from the difference,
$I^{+}-I^{-}$. The XRMS experiments were carried out
at the SIM beamline at the Swiss Light Source using the RESOXS
chamber \cite{jaouen}. The reflected intensities $I^{+}$ and $I^{-}$
were collected for two opposite directions of a magnetic field
applied along the intersection of the superlattice and scattering
planes. The amplitude of the magnetic field available in the chamber
is 0.16 T, which sufficient to field cool the sample into a partially magnetized state with a net magnetization of $\sim$25\% of the saturation value.
\begin{figure}[t]
\includegraphics[height=6cm]{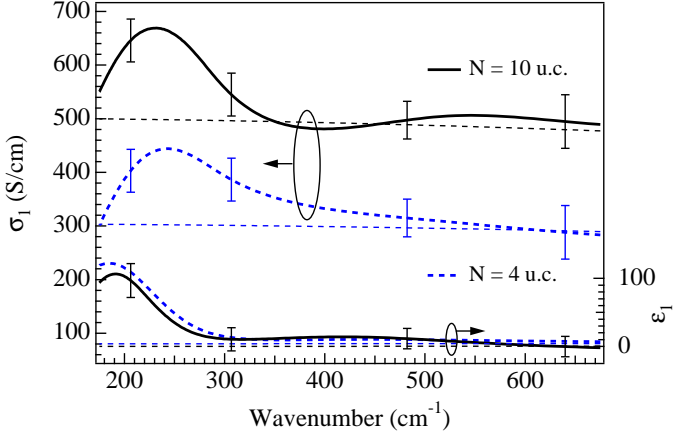}
\caption{Best-fit model functions $\sigma_1 (\omega)$ and $\varepsilon_1(\omega)$
for superlattices with $N =4$ and 10 consecutive CaRuO$_3$ u.c., as obtained from FIR ellipsometry data by inversion of the ellipsometric parameters. The
dashed lines represent the results of Drude fits, as described in the text. The peak at 220 cm$^{-1}$ superposed on the broad Drude response is due to a bond-bending lattice vibration of CaMnO$_3$.}
\end{figure}

\begin{figure}[ht]
\begin{tabular}{c}
        \includegraphics[width=0.4\textwidth]{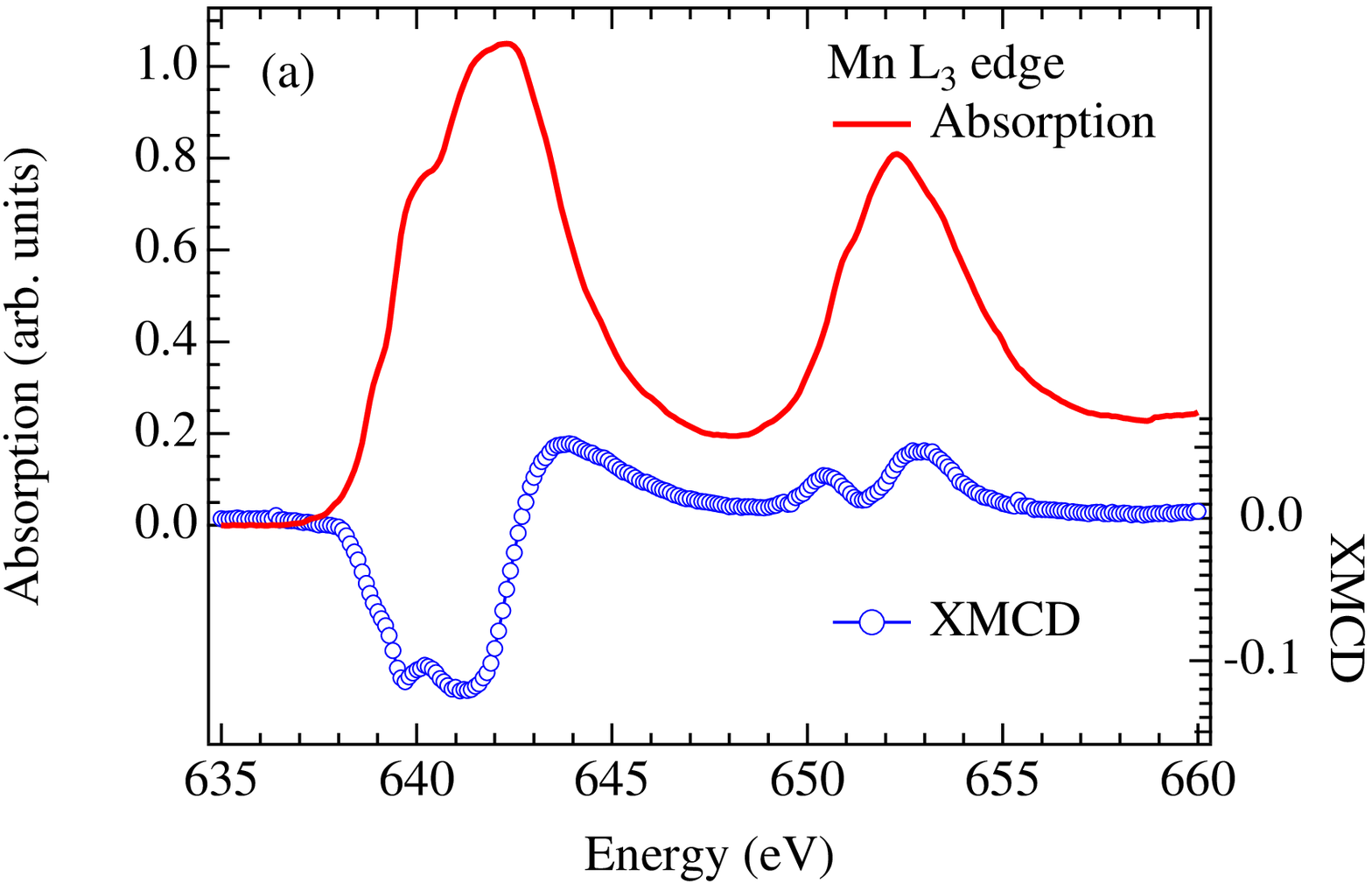} \\
        \includegraphics[width=0.41\textwidth]{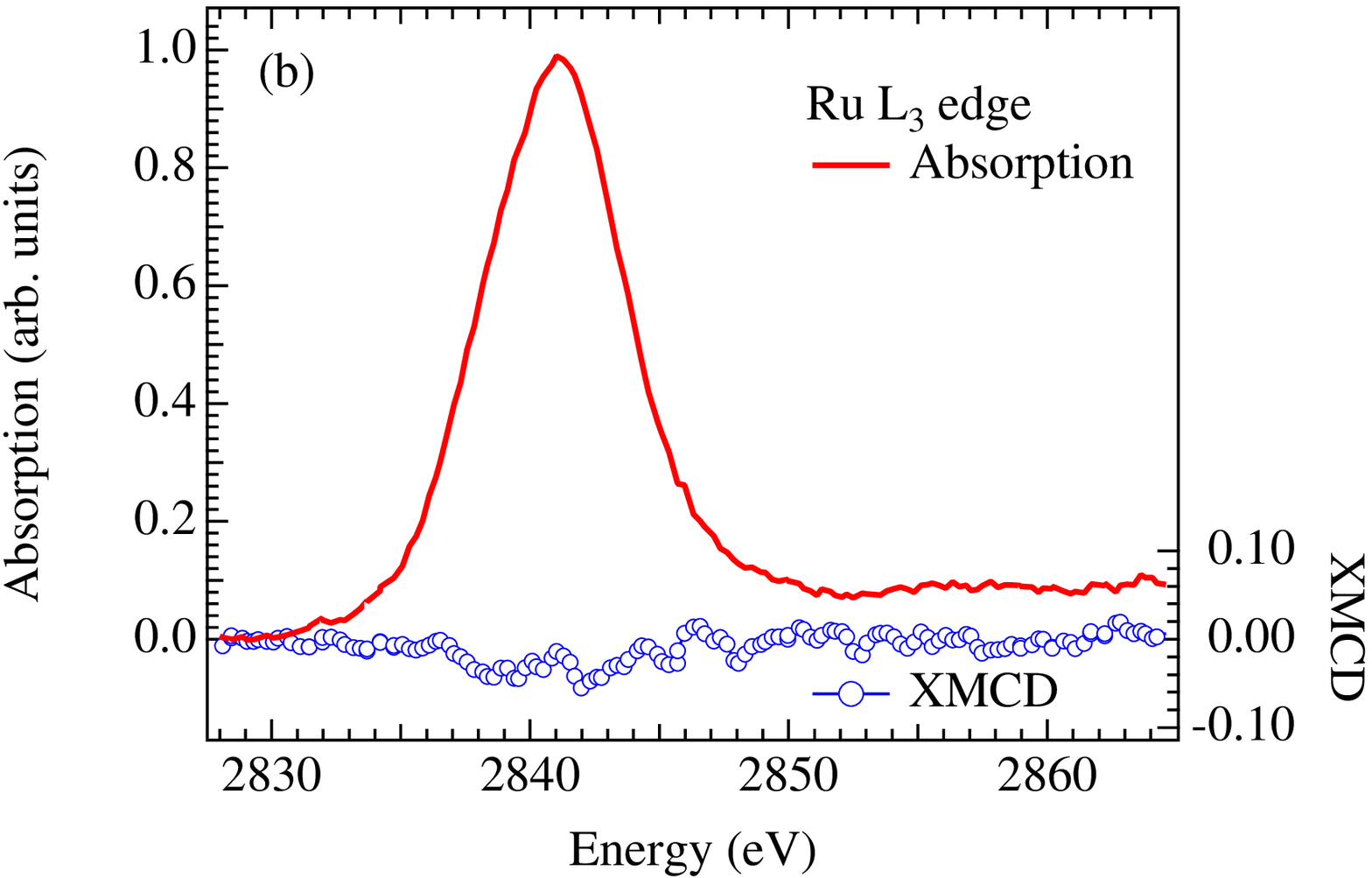} \\
        \includegraphics[width=0.4\textwidth]{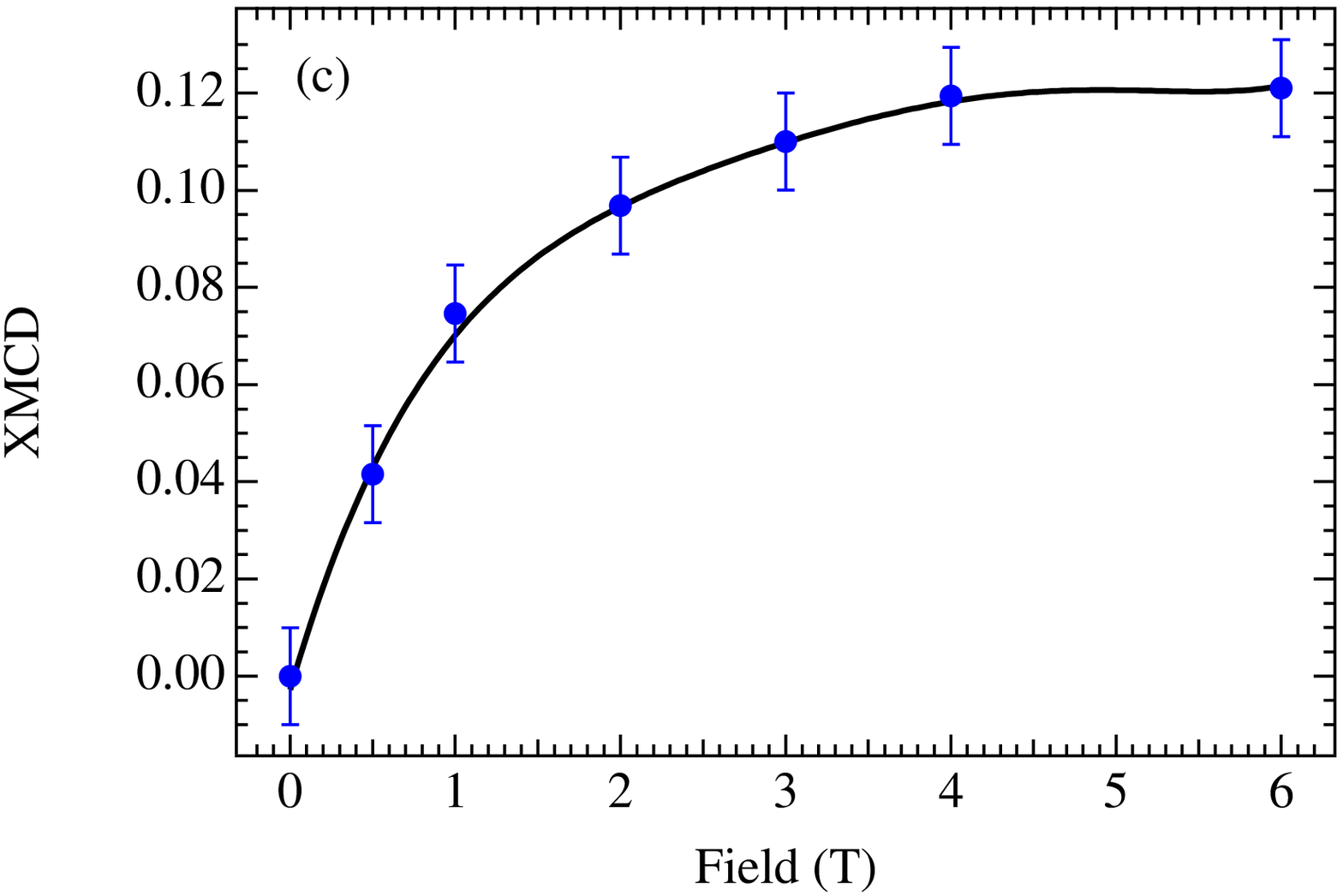} 
\end{tabular}
\caption{(a) Absorption and XMCD measured at the Mn L$_3$ edge at T = 10 K for $N=6$. The XMCD of signal is sizable ($\sim 12\%$) and consistent with rather large magnetic moment. (b) Absorption and XMCD from the Ru L$_3$ edge for the 4 unit cell CaRuO$_3$ sample. Comparison with reference samples indicates that for both the Mn and Ru the valence is close to 4+. The very small apparent Ru XMCD signal is within the systematic error of the measurement.(c) Field dependence of the Mn XMCD measured at 10K showing that the magnetic moment saturates near 4T but shows not sign of remanent magnetization.}
\protect\label{xmcdxrms} 
\end{figure} 

The charge transport properties of the CaRuO$_3$-CaMnO$_3$
superlattices were determined by FIR ellipsometry, which yields the
frequency-dependent complex dielectric function, $\epsilon_1(\omega)
+ i \epsilon_2(\omega)$, without the need for reference measurements
and Kramers-Kronig transformations. In contrast to dc transport
experiments, this method is insensitive to the influence of
substrate-induced steps or strain-induced dislocations on the
current flow through the atomically thin layers, and it avoids
complications arising from the attachment of electrical contacts.
Fig. 1 shows the real parts of the dielectric function,
$\epsilon_1(\omega)$, and the optical conductivity,
$\sigma_1(\omega) = (1/4\pi) \omega \epsilon_2 (\omega)$, of the $N=4$
and 10 superlattices, which were extracted from the ellipsometric
parameters measured at various angles of incidence using a nonlinear
regression analysis \cite{tompkins}. Both data sets are well
described by a broad Drude response with a ratio of scattering rate
and plasma frequency $\sim 0.2-0.3$, which is typical for bulk
single-crystalline CaRuO$_3$ (dashed lines in Fig. 1). (Note that
the parameters in the Drude fit are well constrained, because both
$\sigma_1$ and $\epsilon_1$ are available. ) The effective number of charge carriers per Ru atom
extracted from a sum-rule analysis, $N_{\rm eff}$ = $\frac{2m}{\pi
e^{2} n_{\rm Ru}} \int \sigma _{1}(\omega) d\omega = 0.11\pm 0.03$
(where $m$ is the free-electron mass and $n_{\rm Ru}$ the density of
Ru atoms), is identical for both $N=4$
and 10 samples. It also agrees (within the
measurement error) with the corresponding quantity $N_{\rm
eff}=0.082$ reported for bulk CaRuO$_3$ \cite{kamal,mazin}. These
results indicate that even in the $N=4$ sample the conductivity of
the metallic layers is not significantly disrupted by Ru-Mn
interdiffusion, testifying to the high quality of the interfaces.
They are also consistent with the theoretical prediction that only
$\sim 5$\% of the charge carrier density in the CaRuO$_3$ atomic
layer closest to the interface is transferred across the interface.
A reduction of the Drude weight corresponding to a charge transfer
of this magnitude is within the error of the FIR data.

XAS measurements near the Ru and Mn L$_3$ absorption edges yield
complementary information on the charge transfer across the
interface.
From a comparison of the absorption profiles (solid lines in Fig. 2)
to reference compounds \cite{abbate,saitoh,ruxas3,ruxas2,ruxas1}, we
conclude that both Mn and Ru are close to the 4+ valence state.
Specifically, in the $N=4$ sample (where half of the Ru atoms are at
the interface) a substantial change in the Ru valence state due to
charge transfer across the interface would result in a noticeable
multiplet splitting of the Ru absorption peak due to population of
$e_g$ orbitals, similar to that seen in compounds containing
Ru$^{4+}$-Ru$^{5+}$ mixtures \cite{ruxas3,ruxas2,ruxas1}. From the
absence of this splitting in the data of Fig. 2b we infer an  interfacial Ru valence close to 4+,
consistent with the FIR data and with the predictions of Ref.
\onlinecite{nanda}.

The XMCD signals from the polarized XAS measurements
yield information about the contribution of each constituent
transition metal ion to the net magnetization (see Fig.
2). While the Ru XMCD does not exceed 3\%, a clear XMCD signal of
$\sim 12$\% is observed at the Mn L-edge for both samples in a
magnetic field of 7 T and does not seem to depend on $N$, which is consistent with previous work\cite{takahashi}. Comparing with Ru XMCD references places an upper bound of $\sim0.3 \mu_B$ on the net magnetic moment per Ru atom while the Mn XMCD corresponds to an average moment of 1 $\mu_B$ per Mn atom\cite{terai}, which implies that the net magnetization is dominated by CaMnO$_3$, consistent with the theoretical work of Ref.
\onlinecite{nanda}. The Mn XMCD data also provide two pieces of
evidence supporting the theoretically predicted canted
antiferromagnetic state localized near the interface. First, the
XMCD signal is substantially lower than expected if the layers were
fully ferromagnetic\cite{terai}, and it does not exhibit
significant hysteresis effects. Second, the XMCD for both
superlattices vanishes upon heating above 100-120 K, the N\'{e}el
temperature of bulk CaMnO$_3$, yet control measurements on an
isolated CaMnO$_3$ film grown on SrTiO$_3$ (not shown)
revealed no XMCD signal above background, which implies that the Mn magnetism is due to the CRO/CMO interface and not due to the entire CMO layer.

\begin{figure}[t]
\includegraphics[height=9cm]{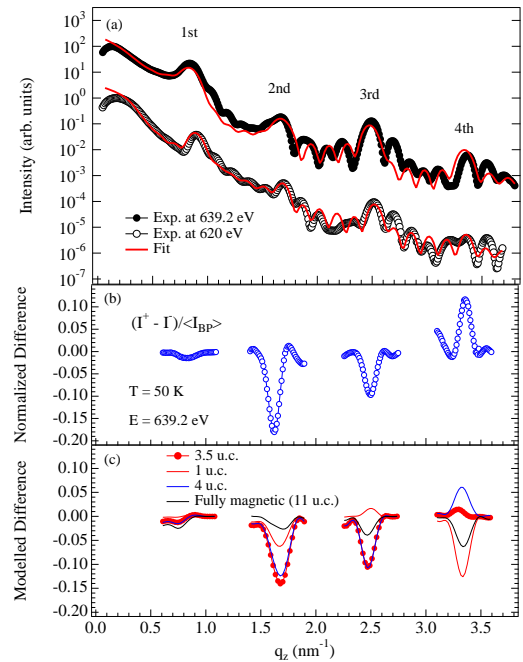}
\caption{(a) Specular reflectivity measured at room temperature with photon energy $E=620$ eV,
below the Mn L$_3$ edge (upper trace) and at 50 K for $E=639.2$ eV on the Mn L$_3$ edge (lower trace).
The lines are the results of fits to a structural model discussed in the text. The Bragg peak positions are indicated. (b) Dichroic differences $I^{+}-I^{-}$ measured at the Bragg peak positions,
normalized to the peak intensity. (c) Calculated normalized difference at the same Bragg
peak positions for different layer thicknesses of the interface magnetized regions in CaMnO$_3$ 
}
\end{figure}

%\begin{figure}[t]
%\includegraphics[height=7cm]{fig4.eps}
%\caption{(upper panel) Dichroic differences $I^{+}-I^{-}$ measured at the Bragg peak positions,
%normalized to the peak intensity. (lower panel) Calculated normalized difference at the same Bragg
%peak positions for different thicknesses, $t^{IF}_{mag}$, of the magnetized regions in CaMnO$_3$ (see text).
%}
%\end{figure}

In order to determine the length scale of the interface-induced
ferromagnetic polarization, we have performed XRMS measurements with
circularly polarized x-rays at two photon energies in the vicinity of the Mn L$_3$ absorption edge. The resonant
reflectivity in the charge channel  ($I^{+}$+$I^{-}$) exhibits a series of Bragg peaks characteristic
of the superlattice periodicity, separated by interference fringes
due the finite thickness of the sample (see Fig. 3(a)). The first photon energy (620 eV) was chosen off resonance allowing a structural analysis not relying on the knowledge of the resonant scattering factor (use of tabulated atomic scattering factor). The second one (639.2 eV, corresponding to the inflexion point of the edge) was chosen to reduce the x-ray absorption while maximizing the real part of the magnetic scattering factor\cite{choi,jwf}.
The
analysis required knowledge of the structural parameters of the
superlattices (thickness, roughness, and density of the layers)
which were derived from a refinement of off-resonant reflectivity
data recorded at 620 eV and confirmed by the refinement of the resonant reflectivity using the resonant scattering factor derived from the XAS measurements\cite{structure}. Although the fits exhibit some discrepancies with the data between the Bragg peaks, the relative intensities of the Bragg peaks and the widths of the intense ones are well described.

The magnetic profile is probed by the the charge-magnetic scattering in the $I^{+}-I^{-}$ signal\cite{choi,jwf}. Fig. 3(b) displays the difference $I^{+}-I^{-}$ measured on top of the
first four Bragg peaks normalized to the corresponding peak intensity
(upper panel), as well as the results of simulations of this
quantity for several models of the magnetic profile (see Fig. 3(c)). For the case of a superlattice, the key point is the $q_z$ evolution due to the cross correlation of the charge and magnetic scattering in the XRMS signal. When the magnetic profile differs from the chemical profile, the sign varies depending on the thickness of the magnetic layer within the charge layer (see Fig. 3(c)).
By considering a simple symmetric step profile for the
magnetization in CaMnO$_3$, composed of equally thick magnetic
layers at the interfaces and a non-magnetic core layer, we can study the evolution of the sign and gain insight into the thickness of the magnetic layer near the interface. 

The simulated and observed intensities show very
good agreement for an interfacial thickness of 1.38 nm (3.5 u.c.).
This model reproduces both the experimentally observed sign sequence
of $I^{+}-I^{-}$ at the Bragg peaks and its relative amplitude. The
absolute magnitude of the normalized difference was calculated with
the constraint that its integral equals the XMCD signal discussed
above (scaled for the lower applied magnetic field, see Fig. 2c). As
shown in Fig. 3(c), the result also agrees closely with the
experimental data. Despite its simplicity, the model therefore
provides an excellent description of the salient features of the
XRMS data.  From this we conclude that the ferromagnetic polarization is not limited to the
immediate vicinity of the interface, as theoretically predicted
\cite{nanda}, but extends further into the CMO layer. Specifically, a model in which the net magnetization
arises entirely from the first CaMnO$_3$ unit cell at the interface
is ruled out by the behavior of $I^{+}-I^{-}$ at the
third- and fourth-order Bragg peaks.

Prior work on lightly La-doped bulk CaMnO$_3$ provides clues to the increased length-scale of the magnetic interface layer and the
origin of the remarkably effective disruption of the
antiferromagnetic order by a small density of electrons.
Specifically, it was shown that this system exhibits a net
magnetization of $\sim 7 \mu_B$ for each electron donated by La
\cite{neumeier}. Based on transport \cite{cohn} and neutron
scattering measurements \cite{argyriou} as well as model
calculations \cite{meskine}, this effect was attributed to the
formation of a magnetization cloud extending over $\sim 7$ Mn sites
around the electron due to the ferromagnetic double exchange
interaction it actuates between neighboring Mn spins, in conjunction
with local lattice distortions. A closely related ``orbital
polaron'' effect has also been investigated in lightly hole-doped
manganates \cite{khaliullin}. A similarly extended cloud of
polarized Mn spins around electrons transferred across the
CaRuO$_3$-CaMnO$_3$ interface would explain the large ferromagnetic
penetration depth we observed. Further work is required to assess
whether the magnetic polarons form a narrow band, or whether they
localize in a charge-ordered state or in the vicinity of defects at
the interface. In any case, our results suggest that magnetic
polarons, which have been extensively investigated in bulk
manganates and other transition metal oxides, strongly influence the
physical properties of oxide interfaces and should therefore be
considered in future theoretical work on this issue.

%In summary, a combination of different spectroscopic probes of
%CaRuO$_3$-CaMnO$_3$ superlattices with atomically sharp interfaces
%has enabled a detailed experimental test of theoretical predictions
%for the interface-induced intermixing of metallicity and magnetic
%order \cite{nanda}. Two key predictions were confirmed: First, the
%density of conduction electrons in the CaRuO$_3$ layers remains
%equal to that of bulk CaRuO$_3$ even if the layers are only four
%unit cells thin, which implies only a small leakage of carriers into
%CaMnO$_3$. Second, the net magnetization arises from canting of Mn
%moments in CaMnO$_3$, with at most a small contribution from Ru ions
%in CaRuO$_3$. However, the length scale for the interface-induced
%ferromagnetic polarization in CaMnO$_3$ exceeds the theoretical
%prediction by more than a factor of three, despite the low density
%of carriers transferred across the interface.

Work at Argonne, including the Advanced Photon Source, is supported by the U.S. Department of Energy, Office of Science, under Contract No.  DE-AC02-06CH11357. We thank
Y.-L. Mathis for the support at ANKA. JC was supported by DOD-ARO
under the Contract No. 0402-17291  and NSF Contract No. DMR-0747808.
HNL was sponsored by the Division of Materials Sciences and
Engineering, U.S. Department of Energy. Work at SLS is supported by the European Commission under the 6th Framework Programme: Strengthening the European Research Area, Research Infrastructures. Contract no¡: RII3-CT-2004-506008.


\begin{thebibliography}{99}

%\bibitem{ahn03} C.H. Ahn et al. Nature {\bf 424}, 1015 (2003).

\bibitem{ahn06} C.H. Ahn {\it et al.} Rev. Mod. Phys. {\bf 78}, 1185 (2006).

%\bibitem{cen} C. Cen {\it et al.},  Nature Mat. {\bf 7}, 298 (2008).

\bibitem{nakagawa} N. Nakagawa, H. Y. Hwang, and D.A. Muller. Nature Mat. {\bf 5}, 204
(2006).

\bibitem{pentcheva} R. Pentcheva and W.E. Pickett, Phys. Rev. B {\bf 78}, 205106 (2008).

\bibitem{tebano} A.  Tebano {\it et al.}, Phys. Rev. Lett. {\bf 100}, 137401
(2008).

\bibitem{chakhalian1} J. Chakhalian {\it et al.},
Science {\bf 318}, 1114 (2007).

\bibitem{chakhalian2} J. Chakhalian {\it et al.}, Nat. Phys. {\bf 2}, 244
(2006).

\bibitem{huijbenrev} M. Huijben {\it et al.}, arXiv:0809.1068 (2008).

\bibitem{nanda} B.R.K. Nanda, S. Satpathy, and M. S. Springborg, Phys. Rev. Lett. {\bf 98},
216804 (2007).

\bibitem{rosch} R.W. Helmes, T.A. Costi, and A. Rosch, Phys. Rev. Lett. {\bf 101}, 066802
(2008).

\bibitem{kamal} S. Kamal {\it et al.}, Phys. Rev. B {\bf 74}, 165115 (2006).

\bibitem{lee} Y.S. Lee {\it et al.}, Rev. B {\bf 66}, 041104 (2002).

\bibitem{mazin} I.I. Mazin and D.J. Singh, Rev. B {\bf 56}, 2556 (1997); {\it ibid.} {\bf 73}, 189903 (2006).

\bibitem{neumeier} J.J. Neumeier and J.L. Cohn, Phys. Rev. B {\bf 61}, 14319 (2000).

\bibitem{takahashi} K.S. Takahashi, M. Kawasaki, and Y. Tokura, Appl. Phys. Lett. {\bf 79}, 1324 (2001).

\bibitem{yamada} H. Yamada {\it et al.}, Appl. Phys. Lett. {\bf 92},
062508 (2008).

\bibitem{bernhard} C. Bernhard {\it et al.}, Thin Solid Films {\bf 455-456}, 143
(2004).

%\bibitem{freeland} J.W. Freeland {\it et al.}, Rev. Sci. Instrum. {\bf 73}, 1408 (2002).

\bibitem{jaouen} N. Jaouen {\it et al.}, J. Synchrotron Rad. {\bf 11}, 353
(2004).

\bibitem{tompkins} H.G. Tompkins and E.A. Irene, eds., {\it Handbook of
Ellipsomety} (Springer, Germany, 2005); A.V. Tikhonravov and M.K. Trubetskov, Optilayer Thin Film Software, \underline{http://www.optilayer.com}

\bibitem{abbate} M. Abbate {\it et al.}, Phys. Rev. B {\bf 46}, 4511
(1992).

\bibitem{saitoh} T. Saitoh {\it et al.}, Phys. Rev. B {\bf 51},
13942 (1995).

\bibitem{ruxas1} G.V.M. Williams, L.Y. Jang, and R.S. Liu, Phys. Rev. B {\bf 65}, 064508
(2002).

\bibitem{ruxas2} R. S. Liu {\it et al.}, Phys. Rev. B {\bf 63}, 212507 (2001).

\bibitem{ruxas3} Z. Hu {\it et al.}, Phys. Rev. B {\bf 61}, 5262
(2000).

\bibitem{terai}
K. Terai et. al. Phys. Rev. B {\bf 77}, 115128 (2008).

\bibitem{choi} Y. Choi {\it et al.}, Phys. Rev. B {\bf 70}, 134420
(2004).

\bibitem{jwf}
J.W. Freeland {\it et al.}, J. Phys. Cond. Mat. {\bf 19}, 315210 (2007).

\bibitem{structure}
The layer parameters derived from x-ray data for the nominally $N = 10$ sample ( $\rm 42.1 \pm 0.8 \AA$ for CaMnO$_3$, $\rm 33.3 \pm 0.6 \AA$ for CaRuO$_3$, corresponding to $N=9$, and an overall roughness of $5 \pm 0.5 \AA$ for both layers probably mostly originating from a low density of substrate steps) deviate slightly from those inferred from RHEED. This was taken into account in the analysis of the XRMS profiles.

\bibitem{cohn} J.L. Cohn, C. Chiorescu, and J.J. Neumeier, Phys. Rev. B {\bf 72}, 024422
(2005); C. Chiorescu, J.L. Cohn, and J.J. Neumeier, {\it ibid.} {\bf
76}, 020404(R) (2007).

\bibitem{argyriou} C.D. Ling {\it et al.}, Phys. Rev. B {\bf 68}, 134440 (2003).

\bibitem{meskine} H. Meskine, T. Saha-Dasgupta, and S. Satpathy, Phys. Rev. Lett. {\bf 92}, 056401 (2004); H. Meskine and S. Satpathy, J. Phys.:
Condens. Matter {\bf 17}, 1889 (2005).

\bibitem{khaliullin} R. Kilian and G. Khaliullin, Phys. Rev. B {\bf 60}, 13458
(1999).



\end{thebibliography}
\end{document}